\documentclass[aps,prd, preprintnumbers,nofootinbib,superscriptaddress]{revtex4}
\usepackage[dvips]{graphicx}
\usepackage{bm,latexsym,amsmath,amssymb,amsfonts,color}
\begin{document}

\title{Semiclassical decay of de Sitter space into black holes \\ with vortex-deformed horizons}

\author{Andr\'es Gomberoff}
\email{andres.gomberoff@umayor.cl}
\affiliation{Facultad de Ciencias, Ingeniería y Tecnología, Universidad Mayor, Avenida Manuel Montt 367, Santiago, Chile.}
\affiliation{Centro Multidisciplinario de F\'isica, Vicerrector\'ia de Investigaci\'on, Universidad Mayor, Camino La Pir\'amide 5750, Santiago, Chile.}
\author{Carla Henr\'iquez-Baez}
\email{carla.henriquez@umayor.cl}
\affiliation{Centro Multidisciplinario de F\'isica, Vicerrector\'ia de Investigaci\'on, Universidad Mayor, Camino La Pir\'amide 5750, Santiago, Chile.}
\author{Aldo Vera}
\email{aldo.vera@umayor.cl}
\affiliation{N\'ucleo de Matem\'atica, F\'isica y Estad\'istica, Universidad Mayor, Avenida Manuel Montt 367, Santiago, Chile.}
\affiliation{Centro Multidisciplinario de F\'isica, Vicerrector\'ia de Investigaci\'on, Universidad Mayor, Camino La Pir\'amide 5750, Santiago, Chile.}

\begin{abstract}
We study the decay of de Sitter space into black holes whose horizons are
dressed by BPS vortices of a $\mathrm{CP}^1$ action. The process is mediated 
by a regular Euclidean instanton obtained as a vortex-deformed
generalization of the Nariai instanton. Its Euclidean geometry has the form
$S^2\times\Sigma$, where $\Sigma$ is a compact surface whose geometry is shaped
by the vortex configuration. The resulting decay rates are controlled by a discrete
topological charge, showing that matter vortices open a new topologically
organized family of decay channels for de Sitter space.
\end{abstract}

\maketitle

\newpage

\section{Introduction}

The study of de Sitter (dS) space is a fundamental task in modern theoretical cosmology and quantum gravity. Observationally, dS space provides the mathematical framework to describe both the early inflationary epoch of the universe \cite{Guth:1980zm} and its current accelerated expansion driven by dark energy \cite{SupernovaSearchTeam:1998fmf}. Understanding the global stability of dS space and characterizing its various non-perturbative decay mechanisms is not merely a technical exercise but a critical requirement to determine the viability and lifetime of the cosmic vacuum.

In this regard, it is well known that dS space is semi-classically unstable to the thermal nucleation of black holes \cite{GINSPARG1983245,Bousso:1996au}. For a small cosmological constant $\Lambda$, this semiclassical decay probability is exponentially suppressed, with the rate $\Gamma$ given by
\begin{equation} \label{rate}
    \Gamma = A \exp\left({-\frac{\pi}{G\Lambda}}\right) \ ,
\end{equation}
where $G$ is Newton's constant, and the prefactor $A$ describes one-loop quantum corrections around the instanton. Other decay channels have been identified when matter fields are present: charged black holes \cite{Mann:1995vb} and dilaton black holes \cite{Bousso:1996pn} are two well-studied examples. A less explored possibility is that the matter sector itself admits topologically nontrivial configurations, giving rise to additional decay channels labeled by discrete charges. The $\mathrm{CP}^1$ non-linear sigma model provides a minimal setting in which this possibility can be studied explicitly.

Recently, in Ref.~\cite{Canfora:2026col} (see also Ref.~\cite{Canfora:2026kwj}), an analytic family of black holes with vortex-deformed horizons was constructed within the full $SU(2)$ non-linear sigma model. Here, we first show that these solutions lie entirely within a $\mathrm{CP}^1$ sector and can be obtained from General Relativity coupled directly to a $\mathrm{CP}^1$ non-linear sigma model.
The $\mathrm{CP}^1$ theory provides the minimal setting for the configurations of interest: its holomorphic sectors carry an integer topological charge and saturate a BPS bound \cite{DAdda:1978vbw,Manton:2004tk,Shifman:2012zz}. 
Although the $\mathrm{CP}^1$ and $SU(2)$ formulations yield the same classical black hole solutions, they are not equivalent at the semiclassical level. They have different negative-mode content around the Euclidean solutions, as we discuss in Sec.~\ref{conclusions}.

The relevance of these black hole solutions to the present problem is twofold. First, their matter configurations carry a discrete topological charge and satisfy BPS equations.\footnote{BPS bounds of this type have recently been studied in flat spacetime in Refs. \cite{Canfora:2023zmt,Canfora:2025qkl,Canfora:2025jqr,Cacciatori:2025irb,Cunha:2026xsl,Canfora:2024mkp,Canfora:2026uxq}.} In the $\mathrm{CP}^1$ description, the zeros and poles of a meromorphic field define vortex cores, around which its phase winds by an integer multiple of $2\pi$. Second, the back-reaction of these multivortex configurations deforms the intrinsic geometry of the black-hole horizon, which loses its spherical symmetry and develops localized ``bumps''. The quantized phase winding around the vortex cores may therefore be viewed as an analog of the Onsager–Feynman circulation condition for superfluid vortices, realized here in a fully back-reacted gravitational setting \cite{onsager1949,feynman1955}.

It is important to highlight that the presence of vortices in black hole physics has been previously explored, particularly from a quantum perspective, where vortices emerge as microscopic excitations \cite{Dvali:2021ofp}. In contrast, the bumpy black holes considered here are fundamentally different: they represent exact, macroscopic, and fully backreacted classical solutions of GR coupled to non-linear matter.\footnote{For other relevant examples of black holes with non-constant horizon curvature, see, for example, Refs. \cite{Novikov:1992gow,Collins:2004ex,Vigeland:2011ji,Emparan:2014pra,Licht:2020odx,Chen:2016rjt,Ferrero:2020twa,Giri:2021xta,Dvali:2021ofp}.}

In this paper, we show that dS space can also decay into this family of bumpy black holes through a vortex-dressed generalization of the Nariai instanton. These dS solutions admit a Nariai limit, corresponding to the maximal mass configuration. The Euclidean continuation of these solutions is smooth and defines a gravitational instanton of the form $S^2\times\Sigma$, where $S^2$ is a sphere of radius $\Lambda^{-1/2}$, and $\Sigma$ is a smooth compact surface of spherical topology, whose geometry and area are determined by the vortex configuration. 

The paper is organized as follows. In Sec. \ref{sectionbh}, we review the bumpy black hole solution reported in Ref. \cite{Canfora:2026col} in the case of a positive cosmological constant. In Sec. \ref{nariai}, we explore the Euclidean section of such a configuration, study its Nariai limit, display the instanton solution, and establish its regularity. In Sec. \ref{action}, we compute the action of the instanton and derive the semiclassical decay rate. Finally, in Sec. \ref{conclusions}, we discuss the physical interpretation of the decay process and comment on open questions and future directions.

\section{The vortex-deformed black hole}
%%%%%%%%%%%%%%%%%%%%%%%%%%%%%%%%%%%%%%%%%%%%%%%%%%%%%%%%
\label{sectionbh}

In this section, we review and discuss the properties of a class of black hole solutions found in Ref. \cite{Canfora:2026col} in a slightly different setting. We consider a $\mathrm{CP}^1=SU(2)/U(1)$ non-linear sigma model minimally coupled to gravity with a positive cosmological constant $\Lambda>0$. The target space $\mathrm{CP}^1$ is diffeomorphic to the two-sphere $S^2$, which can be parameterized using complex stereographic coordinates $(w,\bar{w})$. The action is given by
\begin{equation}
I[g_{\mu\nu},w] = \frac{1}{2\kappa}\int_{\mathcal{M}} d^4x\sqrt{-g}
\left(R - 2\Lambda - \frac{2K}{(1+|w|^2)^2}\,g^{\mu\nu}\partial_{\mu}w\,\partial_{\nu}\bar{w}
\right) \ , \label{I}
\end{equation}
where $R$ is the Ricci scalar, $g$ is the metric determinant, $\kappa=8 \pi G$ is the gravitational coupling, and $K>0$ is a dimensionless coupling constant.

We focus on configurations that extremize the action in Eq. \eqref{I}. It turns out that their metric is similar to the Schwarzschild–dS solution, but with the two-dimensional spheres deformed by vortices produced by the matter fields \cite{Canfora:2026col}.\footnote{In Ref. \cite{Canfora:2026col}, the full $SU(2)$ sigma model was employed. However, the family of solutions considered there belongs to a $SU(2)/U(1)$ sector. Consequently, at the classical level, both formulations yield identical black hole solutions with vortex-deformed horizons. For the semiclassical quantum process analyzed in the present work, the $\mathrm{CP}^1$ framework is the appropriate one.} The spacetime metric reads
\begin{equation}\label{bhm}
ds^2 = -\left(1-\frac{2mG}{r}-\frac{\Lambda}{3}r^2\right) dt^2 + \left(1-\frac{2mG}{r}-\frac{\Lambda}{3}r^2\right)^{-1} dr^2 + r^2 e^{u(x,y)} d\Omega^2 \,,
\end{equation}
where 
\begin{equation}\label{measure}
d\Omega^2 = \frac{4}{(1+|z|^2)^2} \, dz d\bar{z} \,, 
\end{equation}
is the metric of the unit sphere in complex stereographic coordinates $z=x+i y$, and $e^{u(x,y)}$ is a conformal factor that encodes the deformation. We denote by $\Sigma_0$ the round unit sphere and by $\Sigma$ the compact surface endowed with the metric $e^{u(x,y)} d\Omega^2$.

Assuming that the matter fields depend only on the coordinates of the sphere, $w(x,y)$ and $\bar{w}(x,y)$, the equations simplify to
\begin{equation}\label{eom}
\partial_z\partial_{\bar{z}}w - \frac{2\bar{w}}{1+|w|^2}\,\partial_z w \,\partial_{\bar{z}}w = 0 \ , \quad \partial_z\partial_{\bar{z}}\bar{w} - \frac{2w}{1+|w|^2}\,\partial_z \bar{w}\,\partial_{\bar{z}} \bar{w} = 0\,.
\end{equation}

The fields $w(z,\bar{z})$ and $\bar{w}(z,\bar{z})$ define a map $\Sigma_0 \rightarrow S^2$, whose topological charge is given by 
\begin{equation}
Q = \frac{i}{2\pi}\int \frac{dw\wedge d\bar{w}}{(1+|w|^2)^2}
= \frac{1}{\pi}\int \frac{|\partial_z w|^2 - |\partial_{\bar{z}}w|^2}{(1+|w|^2)^2}\,d^2x\,.
\end{equation}
This topological charge measures how many times the field $w:\Sigma_0\to \mathrm{CP}^1$ covers the target sphere $\mathrm{CP}^1$. 

The energy density of the configuration is given by
\begin{equation}
\varepsilon \equiv -\sqrt{-g}\,T^t{}_t = \frac{2K}{\kappa}\frac{|\partial_z w|^2 + |\partial_{\bar{z}}w|^2}{(1+|w|^2)^2}\,.
\end{equation}
Integrating the energy density $\varepsilon$ over $\Sigma_0$ at fixed $r$, and denoting this integral by $E$, one finds
\begin{equation}\label{bound}
E \geq \frac{2\pi K}{\kappa}|Q|\,.
\end{equation}
Note that when $w$ is either holomorphic or anti-holomorphic, the field equations in Eq. \eqref{eom} are automatically satisfied. Assuming that $w$ is holomorphic,
\begin{equation}
\partial_{\bar{z}}\, w = 0\,. \label{BPS}
\end{equation}
the inequality in Eq. \eqref{bound} saturates, yielding $E = 2\pi K|Q|/\kappa$. This is the hallmark of a BPS configuration: a first-order condition that simultaneously solves the second-order field equations and minimizes the energy within each topological sector.
The holomorphic sector describes vortices with $Q>0$ that stretch along the radial direction. The anti-holomorphic sector is completely analogous and describes configurations with $Q<0$, which we refer to as anti-vortices.

If $Q=1$, the map covers $\mathrm{CP}^1$ exactly once, and therefore 
$w$ must pass through every point of it—in particular, through the 
north pole ($w=0$) and the south pole ($w=\infty$). This means that 
$w$ must vanish at some point $z_0\in\Sigma$ and diverge at some 
other point $z_1\in\Sigma$. The simplest holomorphic map with $Q=1$ is
\begin{equation}\label{pair}
w_{\mathrm{pair}}(z)=z\,,
\end{equation}
which has a zero at $z=0$ and a pole at $z=\infty$. We say that there is a vortex of charge $q=+1$ at $z=0$ and a vortex of charge $q=-1$ at $z=\infty$ because the phase of $w$, say at $z=0$, has the quantized winding
\begin{equation}
\frac{1}{2\pi}\oint_{C} d\, \arg w=q \, ,
\end{equation}
where $C$ is a small positively oriented contour around the corresponding core. The charge is simply the order of the zero or pole of $w$ at that point, so that $w=z^q$ corresponds to a pair of vortices of charge $q$ and $-q$. Because $w$ must be single-valued, the charge must be an integer. Note, for consistency, that we may interchange the north and south poles by rotating the sphere using the transformation $z\rightarrow 1/z$. This corresponds to bringing the charge $-q$ at the south pole to the north pole $z=0$.

In the general case, the holomorphic map with vortices of charges 
$q_i$ at positions $z_i$ is
\begin{equation}\label{matter}
w(z) = \prod_i (z-z_i)^{q_i}\,,
\end{equation}
where $q_i\in\mathbb{Z}$. At the south pole $z=\infty$, this formula implies the existence of a charge
\begin{equation}
   q_s= - \sum_i q_i \,.
\end{equation}
That the total charge must vanish stems from the compact nature of the target space, which requires the total number of zeros and poles to balance out when counted with their respective multiplicities. More mathematically, using the argument principle for meromorphic functions on a compact surface, the number of poles and zeros must match. Therefore,
\begin{equation}\label{vanish}
\sum_a  q_a = 0\,,
\end{equation}
where the index $a$ runs over all vortices $q_i$
, including the one with charge $q_s$ at the south pole.

The topological charge of the vortex configuration in Eq. \eqref{matter} is given by
\begin{equation}\label{Qh}
Q = \frac{1}{\pi}\int \frac{|\partial_z w|^2}{(1+|w|^2)^2}\,d^2x = \frac{1}{\pi}\int \partial_z\partial_{\bar{z}}\log(1+|w|^2)\,d^2x = \frac{1}{2}\sum_a |q_a|\,,
\end{equation}
where we have used Eq. \eqref{vanish}.

We now turn to the back-reaction of these vortices on the geometry, which
determines the conformal deformation $u(x,y)$ of the sphere in
Eq. \eqref{bhm}. The remaining Einstein equation is a Liouville equation with a
source fixed by the BPS vortex configuration:
\begin{equation}\label{Liouvilleu}
\partial_z\partial_{\bar{z}}\,u + \frac{2(e^u-1)}{(1+|z|^2)^2} = 
-\frac{K}{(1+|w|^2)^2}\,|\partial_z w|^2\,.
\end{equation}
The source term on the right-hand side comes from the energy-momentum
tensor of the vortex configuration.
Although $w$ has poles, note that the source is regular at the vortex cores. In fact, one can check that, in general, close to $z=z_i$, the source term tends to $-2Kq_i^2|z-z_i|^{2(|q_i|-1)}$. Thus, the metric
deformation $u$ is sourced by a smooth stress tensor of the $\mathrm{CP}^1$ field. 

There is, however, an interesting exception: when two vortices of opposite charges approach the same location $z_0$. Take, for instance, $w=(z-z_0)(z-z_0+d)^{-1}$, with $d\rightarrow0$. Even though $w$ converges to $1$ in the limit, this cannot tend to the vacuum solution $Q=0$, because initially $Q=1$. What happens is that a delta function singularity builds up at $z_0$, forming one-dimensional spikes that pinch the black hole horizon \cite{Canfora:2026col}.

To solve Liouville's equation \eqref{Liouvilleu}, one must fulfill a global condition on the vortex configuration. It is obtained by integrating it over $\Sigma_0$. The first term vanishes since it is the integral of a total derivative over a compact surface. The second term is the subtraction of the measure on $\Sigma_0$ and the one on $\Sigma$, and therefore its integral gives $A_\Sigma-A_{\Sigma_0}$, where $A_\Sigma$ is the area of the surface $\Sigma$, and $A_{\Sigma_0}=4\pi$ is the area of the unit sphere. Finally, the integral on the right-hand side is obtained from Eq. \eqref{Qh}. Putting everything together, we find
\begin{equation}\label{area}
    A_\Sigma = 4\pi\left(1 - \frac{K}{4} \sum_a |q_a|\right)\,.
\end{equation}
The vortices, therefore, reduce the area of the angular surfaces at fixed $r$. This area must be positive; therefore, the following bound is obtained:
\begin{equation}\label{constq}
 \frac{KQ}{2}=\frac{K}{4}\sum_a |q_a| < 1\,.
\end{equation}

In general, Eq. \eqref{Liouvilleu} can be solved provided this inequality is satisfied (see Appendix \ref{App} for the demonstration). It is difficult, however, to provide analytic solutions, with two simple exceptions: the trivial case with no vortices and $u=0$, and the antipodal vortex pair of Eq. \eqref{pair}. It is straightforward to see that, in that case, a solution of \eqref{Liouvilleu} is
\[
e^u=1-\frac{K}{2}\,.
\]
Notably, despite the matter configuration being only axially symmetric, the back-reacted two-geometry $\Sigma$ is round. Its area is reduced by the factor $(1-K/2)$, which is positive because of the condition in Eq. \eqref{constq} that we have already imposed.

Thus, the resulting spacetime coincides with the metric of a Schwarzschild-dS 
black hole with a global monopole (solid angle deficit), although the deficit 
is sourced by BPS vortices in a $\mathrm{CP}^1$ model rather than by a scalar 
triplet. Its causal structure is therefore that of Schwarzschild-dS: 
when $0\leq m<(3G\sqrt{\Lambda})^{-1}$ there are two horizons, the black hole 
horizon $r_+$ and the cosmological horizon $r_{++}>r_+$. 

It is important to note that, even though the matter fields are independent of $r$, which may suggest that the total energy could not be finite, it is indeed finite. The mass can be computed using, for instance, Hamiltonian methods \cite{Gomberoff:2003ea}, and the only difference with the Schwarzschild-dS case is in the integration along the angular variables, which results in the reduced area discussed around Eq.~\eqref{area}. The mass turns out to be 
\begin{equation}
M=\left(1 - \frac{K}{4} \sum_a |q_a|\right) m\,.
\end{equation}

%%%%%%%%%%%%%%%%%%%%%%%%%%%%%%%%%%%%%%%%%%%%%%%%%%%%%%%%
\section{Euclidean continuation and the Nariai limit}
%%%%%%%%%%%%%%%%%%%%%%%%%%%%%%%%%%%%%%%%%%%%%%%%%%%%%%%%
\label{nariai}

In many physical contexts, metastable states may decay either by quantum
tunneling \cite{Coleman:1977py} or by thermal activation
\cite{LANGER1969258, LINDE198137, Linde:1981zj, Affleck:1980ac}. In both
cases, the leading semiclassical contribution to the decay rate is obtained
from Euclidean saddle points of the path integral, namely instantons. These
are solutions of the Wick-rotated field equations and determine the dominant
exponential factor in the transition probability.

In gravitational systems, these saddle points are Euclidean spacetime
geometries. When horizons are present in the Lorentzian sector, the Wick
rotation turns the near-horizon $(r,t)$-plane into a two-dimensional polar
plane: Euclidean time becomes an angular coordinate, the origin of which corresponds
to the horizon. The periodicity of Euclidean time must be chosen so that
conical singularities are avoided at this origin. This period corresponds to
the inverse Hawking temperature of the horizon.

The simplest example in the present context is pure dS space, which has
a cosmological horizon with temperature
$$T_{\mathrm{dS}}=\frac{1}{2\pi}\sqrt{\frac{\Lambda}{3}}\,.$$
Because of this temperature, dS space can decay through the thermally
activated nucleation of maximal-mass Schwarzschild-dS black holes,
characterized by
$$m_{\mathrm{Nariai}}=\frac{1}{3G\sqrt{\Lambda}}\,.$$
In this case, the black-hole and cosmological horizons have the
same area. In Schwarzschild coordinates, taking the limit $m\to m_{\mathrm{Nariai}}$ is subtle because, although the two horizon radii approach the same value in this coordinate system, the proper distance between them remains finite and equal to $\pi\Lambda^{-1/2}$; see Refs. \cite{Nariai:1951, GINSPARG1983245}. The limiting metric is
\begin{equation}\label{nariailorentz}
    ds_{\mathrm{Nariai}}^2 =
\frac{1}{\Lambda}
\left(
-\sin^2\chi\, d\psi^2 + d\chi^2
\right)
+
\frac{1}{\Lambda}\, d\Omega^2\,,
\end{equation}
where $0\leq\chi\leq\pi$. The two horizons are located at
$\chi=0$ and $\chi=\pi$.
After nucleation, the exact Nariai configuration is unstable. Hawking radiation
drives the black hole horizon to shrink while the cosmological horizon expands \cite{Bousso:1996au}.

The semiclassical decay rate $\Gamma$ is computed by considering the Euclidean continuations of each of these solutions. For dS space, this corresponds to a 4-sphere with radius $\sqrt{3}\Lambda^{-1/2}$, a well-known result. Now, performing a Wick rotation, $\psi\to -i\tau$, to the Nariai solution in Eq. \eqref{nariailorentz}, the metric becomes that of two 2-spheres of the same radius $\Lambda^{-1/2}$,
\begin{equation}
    ds_{\mathrm{E-Nariai}}^2 =
\frac{1}{\Lambda}
\left(
\sin^2\chi\, d\tau^2 + d\chi^2
\right)
+ \frac{1}{\Lambda}\, d\Omega^2\,,
\end{equation}
where $0\leq\tau<2\pi$. To take the Nariai limit, one only changes coordinates in the $(r,t)$-plane of the Schwarzschild-dS metric. The same procedure can be applied to the vortex-deformed black hole (see Eq. \eqref{bhm}). Since the $(t,r)$ part of the metric
is the same as in Schwarzschild–de Sitter, the position of the
horizons and the Nariai value of the mass parameter are unchanged. 
The Euclidean metric is now
\begin{equation}
    ds_{\text{E-Vortex}}^2 =
\frac{1}{\Lambda}
\left(
\sin^2\chi\, d\tau^2 + d\chi^2
\right)
+ \frac{1}{\Lambda}\, e^{u(x,y)} d\Omega^2\,.
\end{equation}
The first factor is therefore the same regular two-sphere of radius
$\Lambda^{-1/2}$ as the standard Nariai instanton in empty space. 
The effect of the vortices is entirely contained in the angular surface
$\Sigma$, whose geometry is deformed by the conformal factor $e^{u(x,y)}$.
Although $\Sigma$ has spherical topology, its geometry is not round in general. Still, the
instanton remains smooth because the source in Eq. \eqref{Liouvilleu} is regular
and the surface $\Sigma$ has a positive area, provided the bound in Eq. 
\eqref{constq} is satisfied. The full Euclidean instanton has a geometry $S^2\times\Sigma$. For vanishing vortices, $u=0$ and $\Sigma=S^2$, one recovers the usual
Nariai instanton $S^2\times S^2$.

%%%%%%%%%%%%%%%%%%%%%%%%%%%%%%%%%%%%%%%%%%%%%%%%%%%%%%%%
\section{The decay rate}
%%%%%%%%%%%%%%%%%%%%%%%%%%%%%%%%%%%%%%%%%%%%%%%%%%%%%%%%
\label{action}

The decay rate is calculated by comparing the values of the Euclidean 
actions of the saddles associated with the Euclidean continuations of 
the initial dS configuration and the final vortex-deformed Nariai 
configuration. The semiclassical expression reads
\begin{equation}\label{rateg}
\Gamma = A\exp[-B]\,,
\end{equation}
where the prefactor $A$ describes one-loop corrections, and the 
semiclassical exponential suppression $B$ is
\begin{equation}\label{B}
B=I_E^{\mathrm{vortex}}-I_E^{\mathrm{dS}}\,.
\end{equation}
Here $I_E^{\mathrm{dS}}$ and $I_E^{\mathrm{vortex}}$ are the Euclidean actions 
evaluated on dS space and the vortex-deformed Nariai space, 
respectively. The Euclidean action is obtained from the Lorentzian 
one (see Eq.~\eqref{I}) by the Wick rotation $t\to-i\tau$, and 
setting $I_E=-iI$. One finds
\begin{equation}
I_E[g,w]=\frac{1}{2\kappa}\int d^{4}x\,\sqrt{g}\,
\left(-R+2\Lambda+
\frac{2K}{(1+|w|^2)^2}\,g^{\mu\nu}\partial_\mu w\,\partial_\nu\bar{w}
\right) \ .
\end{equation}
The Euclidean equations of motion take the same form as in the 
Lorentzian case,
\begin{equation}\label{Eeq}
R_{\mu\nu}-\frac12 R g_{\mu\nu}+\Lambda g_{\mu\nu}=\kappa T_{\mu\nu}\,,
\end{equation}
with the energy-momentum tensor $T_{\mu\nu}$ given by
\begin{equation} \label{Tmunu}
T_{\mu\nu} = \frac{K}{\kappa (1+|w|^2)^2}\left[
\partial_\mu w\,\partial_\nu\bar{w}+\partial_\nu w\,\partial_\mu\bar{w}
-g_{\mu\nu}\partial_\alpha w\,\partial^\alpha\bar{w}
\right]\,.
\end{equation}
Taking the trace of Eqs.~\eqref{Eeq} and \eqref{Tmunu} yields
\begin{equation}
R = 4\Lambda - \kappa T\,, \qquad \mbox{with} \qquad
T = -\frac{2K}{\kappa}\frac{g^{\mu\nu}\partial_\mu w\,\partial_\nu\bar{w}}
{(1+|w|^2)^2}\,.
\end{equation}
Substituting this expression for $R$ into the Euclidean action, the 
matter contribution cancels identically, and the on-shell action 
reduces to
\begin{equation}\label{eaction}
I_E^{\text{on-shell}} = -\frac{\Lambda}{\kappa}\,\mathrm{Vol}_4\,,
\end{equation}
where $\mathrm{Vol}_4=\int d^4x\sqrt{g}$ is the four-dimensional 
Euclidean volume of the instanton.

Since the Euclidean geometry factorizes as $S^2\times\Sigma$, its volume is
the product of the area of the two-sphere of radius $\Lambda^{-1/2}$ and the
physical area of $\Sigma$. Since Eq.~\eqref{area} gives the dimensionless area
$A_\Sigma$, the latter is $A_\Sigma/\Lambda$. Thus
$$\mathrm{Vol}_4=\frac{4\pi}{\Lambda}\frac{A_\Sigma}{\Lambda}\,,$$
and using Eq.~\eqref{area}, we obtain
\begin{equation}
    I_E^{\mathrm{vortex}}
    =
    -\frac{2\pi}{G\Lambda}
    \left(1-\frac{K}{4}\sum_a |q_a|\right).
\end{equation}
In particular, in the limit $K\to0$, the Euclidean action smoothly approaches the pure Nariai value, recovering the standard result for the Nariai instanton \cite{GINSPARG1983245,Bousso:1996au}.

These instantons may be interpreted as describing the thermal decay of dS space into vortex-deformed black hole horizons. This process is a generalization of the pair creation of black holes described in Refs. \cite{GINSPARG1983245,Bousso:1996au}, when a $\mathrm{CP}^1$ field is turned on.

To compute the decay rate, we also need the value of the action of the dS instanton describing the initial state, whose volume is that of a 4-sphere. From Eq. \eqref{eaction}, we obtain the standard result:
$$I_E^{\mathrm{dS}}=-\frac{3\pi}{G\Lambda}\,.$$
Substituting the values of $I_E^{\mathrm{vortex}}$ and $I_E^{\mathrm{dS}}$ into Eqs. \eqref{rateg} and \eqref{B}, we obtain
\begin{equation}\label{ratev}
    \Gamma = A\exp\!\left[-\,\frac{\pi}{G\Lambda} \left( 1 + \frac{K}{2}\sum_a |q_a| \right)\right]\,.
\end{equation}
From the above, we can see that the vortex contribution increases the decay exponent and, therefore, further suppresses the decay process. The number of vortices that may be created is bounded by
Eq. \eqref{constq}. As expected, when $q_a=0$, the standard result of Eq. \eqref{rate} is recovered.

At this point, it is important to highlight that the BPS sector contains a nontrivial moduli space that has not been explicitly
integrated over in the present calculation. For fixed vortex charges $q_a$, the
vortex positions on $\Sigma$ are continuous moduli. Different points in this
moduli space correspond to different local horizon geometries because the
source in Eq. \eqref{Liouvilleu} is redistributed over the angular surface. The
on-shell action, however, is insensitive to these moduli and depends only on the topological charge $Q$ defined in Eq. \eqref{Qh}. Consequently, the moduli do not affect the leading exponential factor in the decay rate but should contribute to the one-loop prefactor $A$, which we leave for future work.

%%%%%%%%%%%%%%%%%%%%%%%%%%%%%%%%%%%%%%%%
\section{Conclusions and outlook}
\label{conclusions}
%%%%%%%%%%%%%%%%%%%%%%%%%%%%%%%%%%%%%%%%

In this paper, we have shown that de Sitter space can semiclassically decay into a family of bumpy black holes whose horizons are dressed by BPS vortices of the $\mathrm{CP}^1$ model. This decay process is mediated by a vortex-deformed generalization of the Nariai instanton, whose Euclidean section is a smooth manifold of the form $S^2\times\Sigma$, where $\Sigma$ is a compact surface representing a non-round deformation of the two-sphere. The resulting semiclassical decay rate, given by Eq. \eqref{ratev}, reveals that the nucleation of these vortex-dressed configurations is further suppressed compared to the standard Nariai case, with the additional suppression growing monotonically with the topological charge $Q$ defined in Eq. \eqref{Qh}.
The interpretation of the Euclidean solution as a decay instanton
requires the presence of a single negative mode in the spectrum of
the quadratic perturbation operator around it. In the standard Nariai 
case, Ginsparg and Perry explicitly identified such a mode \cite{GINSPARG1983245}. 
The present case is richer, and one should proceed with care.

The fluctuation spectrum splits into a matter sector and a gravitational 
sector, with possible mixing between them. For the matter sector, the 
argument is clear: the $\mathrm{CP}^1$ field $w$ satisfies the BPS 
condition $\partial_{\bar{z}}w=0$, which is precisely the condition 
that saturates the energy bound
\begin{equation}
E \geq \frac{2\pi K}{\kappa}|Q|\,,
\end{equation}
meaning that $w$ minimizes the matter action within its topological 
sector. Any deformation of the field cannot decrease the energy. The zero modes include variations of the vortex positions on $\Sigma_0$, which leave the energy unchanged. These modes contribute to the one-loop prefactor $A$ but not to the leading exponential suppression. The matter sector therefore contributes no negative modes.

Classically, the configurations presented here are solutions of both the $SU(2)$ and the $\mathrm{CP}^1$ non-linear sigma models. However, their embedding in the full $SU(2)$ theory introduces additional negative directions in the Euclidean matter action. Thus, the Euclidean solutions constructed here cannot be directly interpreted as decay instantons of the full $SU(2)$ theory.

The gravitational sector, and in particular the mixing between metric 
and matter fluctuations, is more subtle. The vortex-deformed instanton 
reduces continuously to the standard Nariai solution when $K\to0$, 
so the Ginsparg-Perry negative mode is present at $K=0$ and, by 
continuity, it should persist for small $K$. Whether additional negative modes 
can appear as $K$ is increased—either from the gravitational sector 
or from matter-gravity mixing—is an open question that would require 
a complete spectral analysis of the coupled system. We expect, however, 
that for $K$ sufficiently small, no such crossing occurs, and the 
instanton retains exactly one negative mode. A definitive answer to 
this question would be a worthwhile target for future work and may 
reveal new features of the instanton landscape of de Sitter space.

The present results suggest that $\mathrm{CP}^1$ matter, if present 
during an early de Sitter phase, may open additional decay channels 
with distinct suppression factors controlled by the coupling $K$ and 
the topological charge $Q$. Finally, it would be
worthwhile to extend the analysis to charged and dilatonic black
holes in Refs. \cite{Mann:1995vb,Bousso:1996pn}, where the interplay
between gauge fields and non-linear models may yield a richer
instanton landscape.

\begin{acknowledgments}
The authors would like to thank Fabrizio Canfora for many illuminating discussions and insightful suggestions throughout the development of this work. C. H. appreciates the support of FONDECYT postdoctoral Grant No. 3240632. A. V. has been funded by FONDECYT Iniciaci\'on No. 11261883.
\end{acknowledgments}

\appendix

\section{Existence of solutions to the Liouville equation} \label{App}

In this Appendix, we show that the Liouville equation defined in Eq. \eqref{Liouvilleu} admits solutions. Let us first rewrite it in the form
\begin{equation}\label{Liouvilleu2}
\Delta_{S^2}u+2(e^u-1)=
-K\frac{(1+|z|^2)^2}{(1+|w|^2)^2}\,|\partial_z w|^2\,,
\end{equation}
where $\Delta_{S^2}$ is the Laplace--Beltrami operator on the unit sphere.
We begin with the well-known result that the Poisson equation on the two-sphere,
\begin{equation}\label{poisson}
\Delta_{S^2}\Psi=\rho\,,
\end{equation}
admits a solution, unique up to an additive constant, provided that the source $\rho$ is smooth and has a vanishing average over the sphere:
\[
\frac{1}{4\pi}\int_{S^2}d\Omega\, \rho=0 \ ,
\]
where $d\Omega$ is the measure corresponding to the metric in Eq. \eqref{measure},
\[
d\Omega=\frac{4}{(1+|z|^2)^2}\,d^2x \ .
\]
The necessity of this condition follows immediately by integrating Eq.~\eqref{poisson} over the sphere. The full result can be found in Ref.~\cite{Aubin1998}, Theorem 4.7.

We now use this theorem to find a solution, $\Psi$, of Eq.~\eqref{poisson}, with
\[
\rho=
-K\frac{(1+|z|^2)^2}{(1+|w|^2)^2}\,|\partial_z w|^2
+KQ\,.
\]
This source is smooth and has a vanishing average as a consequence of Eq.~\eqref{Qh}. Therefore, a smooth solution $\Psi$ exists. We now define $u=v+\Psi$ and substitute this expression into Eq.~\eqref{Liouvilleu2}. We obtain
\begin{equation}\label{liouville3}
\Delta_{S^2}v=c-he^v\, ,
\end{equation}
where $c=2-KQ$ and $h=2e^\Psi$. Eq. \eqref{liouville3} has the form of the Liouville equation on a compact manifold considered in Theorem 7.2 of Ref.~\cite{Kazdan1974}. There, it is shown that when $h$ is positive somewhere in the manifold, a solution exists for $0<c<c_+(h)$, where $c_+(h)$ is a constant that depends on $h$. For the 2-sphere, the results quoted in that reference imply that $c_+(h)>2$.

In the present case, $h=2e^\Psi$ is positive everywhere. Moreover, Eq.~\eqref{constq} implies
\[
0<c=2-KQ<2\,.
\]
It follows that $0<c<c_+(h)$, and therefore Eq.~\eqref{liouville3} admits a solution. Consequently, the original Liouville equation in Eq. \eqref{Liouvilleu2}, and hence Eq.~\eqref{Liouvilleu}, also admits a solution.

\bibliography{apssamp}

\end{document}